# Tunable atomically enhanced moiré Berry curvatures in twisted triple bilayer graphene

Konstantin Davydov[1], Ziyan Zhu[2], Noah Friedman[1], Ethan Gramowski[1], Yaotian Li[1], Jack Tavakley[1], Kenji Watanabe[3], Takashi Taniguchi[4], Mitchell Luskin[5], Efthimios Kaxiras[6,7], Ke Wang[1*]

[1]*School of Physics and Astronomy, University of Minnesota, Minneapolis, Minnesota 55455, USA*
[2]*Stanford Institute for Materials and Energy Sciences, SLAC National Accelerator Laboratory, Menlo Park, California 94025, USA*
[3] *Research Center for Electronic and Optical Materials, National Institute for Materials Science, 1-1 Namiki, Tsukuba 305-0044, Japan*
[4] *Research Center for Materials Nanoarchitectonics, National Institute for Materials Science, 1-1 Namiki, Tsukuba 305-0044, Japan*
[5]*School of Mathematics, University of Minnesota, Minneapolis, Minnesota 55455, USA*
[6]*Department of Physics, Harvard University, Cambridge, Massachusetts 02138, USA*
[7]*John A. Paulson School of Engineering and Applied Sciences, Harvard University, Cambridge, Massachusetts 02138, USA*
**Corresponding author. Email: kewang@umn.edu*

**We report a twisted triple bilayer graphene platform consisting of three units of Bernal bilayer graphene consecutively twisted at 1.49° and 1.68°. We demonstrate the atomic reconstruction between the two competing moiré superlattices strongly enhances the Berry curvature of each moiré band insulator state, characterized by measured strong nonlocal valley Hall effect that sensitively depends on the inter-moiré competition strength, tunable by manipulating the out-of-plane carrier distribution. Our study sheds light on the microscopic mechanism of atomic and electronic reconstruction in twisted multilayer systems, by systematically investigating transport signatures of moiré Berry curvature and its enhancement from moiré-of-moiré lattice reconstruction. We show that the reconstructed electronic band can be versatilely tuned by electrostatics, providing an approach toward engineering the band structure and its topology for a quantum material platform with designer electrical and optical properties.**

The interference between two slightly rotated atomic lattices, which occurs at the homo and heterointerface of atomically thin materials stacked with small twist angles, gives rise to a moiré superlattice and flat bands where electronic interactions are enhanced. Novel material properties have been reported in such structures, including ferromagnetism [1], the quantum anomalous Hall effect [2–5], quantum spin Hall effect [6], Mott insulators [7–9] and superconductivity [10,11].

Beyond twisted bilayers, the addition of a third 2D layer further enriches the emergent physics [12]. When the third layer aligns with the first layer, the original moiré superlattice is atomically stabilized with enhanced correlated insulator and superconducting states [13–15]. When the third layer is misaligned with the existing layers, two separate moiré superlattices coexist [16–19]. The beating between the two moiré superlattices gives rise to higher order superlattices [20,21], referred to as moiré-of-moiré (MoM) superlattices. A wide range of competing length scales coexist, giving rise to an abundance of new electronic properties [22–24], including an inter-moiré Hofstadter butterfly [19], quasicrystal states [17], correlated charge density wave insulators [23] and signatures of correlated insulating and superconducting states at an extremely low carrier density (~ $10^{10}$ $cm^{-2}$) [25].

To understand how competing atomic structures alter the resulting electronic band in a moiré-of-moiré superlattice, we fabricated a twisted triple bilayer graphene (tTBLG) device consisting of three units of Bernal bilayer graphene (BLG) consecutively twisted at 1.49° and 1.68° [see Fig. 1(a) and Supplemental Material [26]: Section S1]. The choice of different twist angles ensures that the sizes of two coexisting moiré superlattices are sufficiently different [Figs. 1(b)–1(c)], so that the transport signatures from each moiré lattice can be isolated and traced back to insulator states at different carrier densities, and the interaction between the moiré superlattices can be examined via emergent higher order transport features at the carrier density in between full fillings of the two individual moiré patterns. The choice of BLG as the basic unit of stacking allows for characterization of atomically induced band reconstruction by tracking the Berry curvature [27,28]. Berry curvature hot spots were previously reported at the band edges of a parabolic like band [at the charge neutrality point (CNP) of BLG or band insulator (BI) states of a moiré lattice] [29–34], and can be sensitively enhanced by atomic strain/reconstruction from aligned hBN that further breaks the inversion symmetry [33,35–38].

Figure 1(d) shows the calculated band structure of tTBLG at the twist angles of 1.49° and 1.68°. The low energy bands are parabolic instead of linear as in twisted bilayer or trilayer graphene, with the potential of hosting a Berry curvature hot spot at the band edges of moiré and MoM bands. Figure 1(e) shows the measured at applied top, $V_{TG}$, and back gate voltages, $V_{BG}$, four-probe longitudinal local resistance $R_L = R_{xx}$ (Region 1 of the tTBLG device, between Contacts 4 and 3) as a function of the applied displacement field $D$ (having the positive direction toward the graphite gate), and carrier density $n$ (see Supplemental Material [26]: Section S2 for $D$ and $n$ calculations). The top (bottom) axis is given in terms of filling factors for top (bottom) moiré superlattices $\nu_T = 4n/n_T$ ($\nu_B = 4n/n_B$), where $n_T$ ($n_B$) correspond to 4 charge carriers per top (bottom) moiré unit cell. Signature fillings from the top (bottom) moiré patterns are highlighted by yellow (purple) lines, where local resistance peaks are expected at integer fillings, including band insulator states at full fillings (solid lines), and correlated insulator states at half fillings (dashed lines) [23].

At $\nu_T = –4(+4)$, the hole-type (electron-type) charge carries are localized in the top moiré superlattice by a positive (negative) $D$. The tTBLG in its entirety therefore exhibits insulating behavior, confirmed by the temperature dependence [Fig. 3(a), see also Supplemental Material [26]: Section S4], with an estimated thermal activation gap of ~6 meV for $\nu_T = 4$ and ~3 meV for $\nu_T = –4$, respectively.

In tTBLG, we show that the coexistence of two moiré superlattices and their interaction further enhances the Berry curvature of the moiré band insulator states, and can be sensitively manipulated by tuning the inter-moiré interaction strength. The Berry curvature is characterized with the well-established “nonlocal” measurement at $B = 0$ T [30–41]. The valley-specific anomalous velocity drives the ballistic carrier trajectory astray (more so than conventional current diffusion [28,30–38,40,41]), resulting in its detection at voltage probes sufficiently away from the current path.

We apply current $I_{73}$ between Contact 7 and 3 of the tTBLG device [Figs. 1(b) and 2(a)], and measure voltage $V_{64}$ between Contact 6 and 4 (as a result of transverse valley-polarized current) away from the current excitation (nonlocal) to study valley Hall effect (VHE) [30–38,40–43]. The nonlocal resistance $R_{\mathrm{NL}} = V_{64}/I_{73}$ as a function of the carrier density and $D$ field is plotted in Fig. 2(b). Nonlocal resistance peaks are found near $\nu_{\mathrm{T}} = \pm 4$ and $D < 0$ ($D > 0$), satisfying prerequisites for observing a moiré Berry curvature hot spot: (1) Fermi level is near the edge of the band insulator gap, (2) the bottom BLG is depleted of charge carriers (and thus does not contribute to the transport). When both conditions are met, the tTBLG overall is insulating with its Berry curvature well-defined. Near the Berry curvature hot spot, the interlayer screening is negligible, and the displacement field can be approximated as uniform across all 3 layers. Similarly, a nonlocal resistance peak is found near $\nu_{\mathrm{B}} = +4$ and $D > 0$, near the band insulator gap of the bottom moiré while the top BLG is depleted of electron-type carriers. Additionally, substantial nonlocal resistance is observed at CNP at $D > 0$, arising from the Berry curvature hot spot near the displacement-field-induced BLG bandgap [30,31], when all three BLG layers are depleted of charge carriers. The absence of an expected nonlocal resistance peak near $\nu_{\mathrm{B}} = -4$, and CNP at $D < 0$ can be explained by (1) broken electron-hole symmetry in tTBLG; (2) top BLG is not in-gap for $D < 0$ (consistent with a smaller measured $R_{xx}$) making such states less insulating or out of the measurement range.

These resistance peaks arise from the ballistic valley current driven by the Berry curvature, instead of stray currents due to diffusive Ohmic transport, supported by several separate experimental observations. First, the longitudinal resistance peaks [Fig. 1(e)] corresponding to the correlated insulator states are completely absent in the nonlocal measurement [Fig. 2(b)], confirming the origin of nonlocal resistance peaks as curved ballistic transport due to the Berry curvature, instead of a simple diffusive transport at resistive insulating states. Moreover, following the previously-established methods [30–32,34,35,38], we measure the nonlocal signal $R_{NL}$ at voltage probes with varying distance $L$ to the applied current (see Supplemental Material [26]: Section S7). The $R_{NL}$ as a function of $L$ is fitted by $\exp(-L/\xi)$, giving a decay length $\xi = 4\pm2$ μm, a change much slower than the decay due to diffusive Ohmic transport, and consistent with the inter-valley scattering length.

In addition, the trivial Ohmic contribution ($R_{NLO}$) from the stray currents can be estimated from the measured local ($R_L = R_{xx}$) resistance by $R_{NLO} = R_L(w/L)\exp(-L/\lambda)$, where $\lambda = w/\pi$; $w$ = 0.9 μm, $L$ = 1.6 μm are the width and length of the channel. Following the previously well-established methods [30–32,34,35,38], the estimated Ohmic contribution can then be compared to the measured nonlocal resistance to understand the origin of measured $R_{NL}$. As an example [Fig. 2(c)], at $\nu_T = -4$ and $D/\varepsilon_0 = +0.45$ V/nm, the measured nonlocal resistance $R_{NL}$ [scanned across the white line denoted in Fig. 2(b)] is over 100 times larger than expected contribution from the diffusive Ohmic transport $R_{NLO}$ (see Supplemental Material [26]: Section S3 for other band insulator states). The Ohmic contribution is therefore marginal, and the nonlocal resistance is primarily attributed to the Berry curvature.

Last but not least, the power law scaling between the measured local and nonlocal resistances $R_{NL} \sim R_L^{\gamma}$ have been previously used to determine the nature of the nonlocal signal [30–35,37,38,41,42]. Ohmic behavior is linear in nature ($\gamma = 1$), while nonlocal signal caused by the Berry curvature is expected to be nonlinear, with either $\gamma > 1$ or $\gamma < 1$ depending on the saturation of ballistic valley current channel [42]. In contrast to the previously reported Berry curvature hot spot in BLG, twisted bilayer graphene (tBLG) and twisted double bilayer graphene (tDBLG) [30–34,38], the moiré Berry curvature hot spot in tTBLG shows an abrupt change across the band insulator gap. To show this, the measured nonlocal resistance near $\nu_T = -4$ [Fig. 2(c)] is categorized into electron-type (blue) and hole-type (red). The $R_{NL}$ versus $R_L$ [Fig. 2(d)] is then separately plotted for each category of carrier type, and individually fitted against $R_{NL} \sim R_L^{\gamma}$. Both show a good fit to a nonlinear power law dependence, suggesting a Berry curvature hot spot as the underlying mechanism. The quantitative difference in the extracted power $\gamma$ suggests that the Berry curvature hot spot shows a strong electron-hole asymmetry near $\nu_T = -4$, with the Berry curvature hot spot being stronger near the electron-type band edges of the $\nu_T = -4$, leading to a stronger valley Hall current [Fig. 2(e), solid arrows] saturating the maximum current capacity of ballistic conductance channels with $\gamma < 1$. The reverse effect, a weaker Berry curvature hot spot and valley Hall current, correspond to hole-type band edges, with $\gamma > 1$ [see Fig. 2(e), hollow black arrows] [42].

We show that the observed strong Berry curvature hot spot at $\nu_T = -4$ of the top moiré pattern is enhanced by the atomic reconstruction and electronic modification with the bottom moiré pattern, which further breaks the inversion symmetry and alters the band curvature. This can be demonstrated by measuring the power law $\gamma$ [Fig. 2(f)], while tuning the strength of inter-moiré interaction with $D$ [19]. As the $D$ field increases, the electron-type carrier is less localized in the top moiré pattern while its interaction with bottom moiré is enhanced leading to the $\nu_T = -4$ band insulator gaps whose Berry curvature is strongly reinforced by the presence of the bottom moiré pattern (the reverse effect corresponds to decreasing $D$ field, with more localized carriers in the top moiré pattern and suppressed interaction with the bottom moiré pattern). As a result, the power law $\gamma$ is observed to be continuously tunable by the field value ($D$) over one order of magnitude.

At small $D$, when the inter-moiré interaction is weakest, the power law $\gamma = 3$ is similar to the previously reported valley Hall effect without an atomic reconstruction from aligned hBN [33,35,37,38]. At large $D$, when the inter-moiré enhancement is strongest, the record-low $\gamma = 0.5$ suggests an unusually large valley Hall current driven by an extraordinarily-strong Berry curvature, oversaturating the available ballistic channels, more so than atomically-enhanced valley Hall effect previously reported in hBN-aligned systems [35,37,38]. The displacement field dependence is also consistent with the Berry curvature hot spot belonging to the band insulator states of the top moiré superlattice instead of the bottom bilayer graphene, with the latter characterized to yield a weaker valley Hall signal at charge neutrality point $n = 0$.

In contrast, $\gamma = 3$ is observed for the hole-type carriers near $\nu_T = -4$, largely regardless of the $D$ field applied. This suggests that the inter-moiré atomic reconstruction has much weaker implication to the hole-like band edges, possibly due to a large electron-hole asymmetry in band dispersion or different ground state atomic orbitals for electrons and holes.

To extract the characteristic energy scale of the moiré Berry curvature hot spot and to further confirm their microscopic physics pictures, we measure the temperature dependence (Figure 3) of the nonlocal transport. Figure 3(a) plots the measured local resistance at the peak (black) and half peak (red: hole-like carriers; blue: electron-like carriers) as a function of temperature. At temperatures $T > 10$ K, the charge carriers across the $\nu_T = -4$ band insulator gap are thermally activated, resulting in a measured local resistance that decreases with increasing temperature. By fitting the temperature dependence at $T > 10$ K with the Arrhenius law: $R_L \sim \exp\left(\frac{\Delta_L}{2k_B T}\right)$ [Fig. 3(b), dashed], the size of the band-insulator gap is extracted to be on the order of meV [Fig. 4(b), diamonds], consistent with the calculated band structure. The temperature dependences of the peak (black) and half-peak (red, blue) local resistance [Figs. 3(a)–3(b)] give similar band gap sizes [Fig.

4(b), diamonds with the respective colors], indicating thermal activation of carriers at partial filling of the conductance and valence band edges near $\nu_T = -4$.

The nonlocal resistance $R_{NL}$ near the Berry curvature hot spot at $\nu_T = -4$ is also measured at varying temperature [Fig. 3(c)], from 10 mK to 20 K with an equal step size of $\Delta T \sim 2$ K. The resistance peak shows a weak temperature dependence below 10 K (with the curves for $T < 10$ K largely overlapping with each other), while drastically diminishes as $T$ rises above 10 K. The change in temperature dependence can be seen more evidently when plotting $R_{NL}$ as a function of temperature [Fig. 3(d)], measured at the peak (black) and half-height on the hole-like (red) and electron-like (blue). The three sets of data all exhibit a strong (weak) $T$-dependence above (below) $T \sim 10$ K, with fitting to $\exp\left(\frac{\Delta_{NL}}{2k_B T}\right)$ above 10 K yielding a similar thermal activation gap $\Delta_{NL}$. This is consistent with the nonlocal signal (peak or half-height) being expected only when the Fermi level is near mid gap or at the band edges of the band insulator states. The displacement field dependence of the extracted $\Delta_{NL}$ [Fig. 4(b), triangles] is also consistent with that of $\gamma$ [Fig. 2(f)], with a larger $\Delta_{NL}$ being observed at higher $D$ fields when stronger inter-moiré interaction further enhances the Berry curvature hot spot.

At all values of the $D$ field, $\Delta_{NL}$ is consistently larger than the thermal activation gaps $\Delta_L$ at $\nu_T = -4$ [Fig. 4(b)]. The two energy scales near the band insulator states are illustrated in Fig. 4(a), each determining different aspects of the transport behavior near the Berry curvature hot spot: (1) the band insulator gap $\Delta_L = E_e - E_h$, defined as the difference between electron-like band minima $E_e$ and hole-like band maxima $E_h$, the conventional definition of a band gap; this gap determines the conventional transport such as $R_{xx}$. (2) The nonlocal gap $\Delta_{NL} = E_e^* - E_h^*$, defined as the difference between the $E_e^*$ and $E_h^*$, the energy in electron-like and hole-like bands, beyond which ($E > E_e^*$ and $E < E_h^*$) the Berry curvature becomes zero. This gap affects the anomalous transport aided by the Berry curvature, such as $R_{NL}$.

The observed temperature dependence is consistent in each case with the Berry curvature hot spot at $\nu_T = -4$, with a $\Delta_L = 3$ meV and $\Delta_{NL} = 10$-15 meV. At temperature below 10 K (or $\Delta_{NL}/10$), additional carriers may be thermally excited across $\Delta_L$, but they reside at band edges well within the Berry curvature hot spot (between $E_e^*$ and $E_h^*$), equally capable of carrying valley Hall current that saturates the ballistic conductance channel. At temperature above 10 K (or ~ $\Delta_{NL}/10$), a significant portion of the carriers in the applied current start to fall energetically outside the Berry curvature hot spot, and thus stop experiencing the anomalous velocity, no longer contributing to the nonlocal signal, hence the change of temperature dependence [Figs. 3(c)–3(d)] of $R_{NL}$ at $T \sim 10$ K, and a significantly larger $\Delta_{NL}$ [Fig. 4(b)] that more sensitively depends on the $D$ field applied compared to $\Delta_L$. This implies that the enhancement of the Berry curvature hot spot (characterized by $\Delta_{NL}$) is more attributed to the inter-moiré interaction than the evolution of the band structure itself (characterized by $\Delta_L$).

Figure 4(c) presents a log-log plot of $R_{NL}$ vs $R_L$ where each data point is taken at a different temperature $T$. Following the previously-established approaches [31–34], we observe distinct power law behaviors at temperature range above and below $T \sim 10$ K, consistent with our physics picture [Fig 4(a)]. At low temperature range below $T \sim 10$ K [top left part of Fig. 4(c), $R_{NL}$ remains constant while $R_L$ is rapidly decreasing with $T$ (with power law $\gamma < 0.5$)]. In this regime, the ballistic valley conduction channel is saturated, and thermally activated additional carriers are capable of contributing to local transport only. This behavior persists up to critical temperatures of $T_0 \sim 5$ K ($T_0 \sim 8$ K) for the hole- (electron-) like carriers and up to $T_0 \sim 7$ K for the resistance peak, consistent with the larger Berry curvature hot spot at the electron side. At higher temperature ($T > 10$ K), charge carriers are thermally excited away from the Berry curvature hot spot near the band edges. The weaker Berry curvature they experience makes the ballistic valley conducting channel unsaturated, resulting in an observed power law of $\gamma \sim 3$-4 for both charge carrier types [Fig. 4(c)]. This is consistent with the ratio of the previously extracted characteristic energy scales $\Delta_{NL}$, $\Delta_{L}$ [Fig. 4(b)] and the theoretical values [42] of $\gamma$ for an unsaturated valley conductance channel.

In conclusion, we studied a twisted triple bilayer graphene device with consecutive twist angles of 1.49° and 1.68°, in which atomic and electronic reconstruction between two coexisting moiré superlattices gives rise to significantly enhanced moiré Berry curvature hot spots. The latter result in amplified valley currents near moiré band insulator states, characterized by the measured strong nonlocal valley Hall effect, which depends sensitively on electrostatically tuned inter-moiré competition strength, supporting several orthogonal sets of experimental observations. Our study sheds new light on the microscopic nature of atomic and electronic properties stemming from the interplay and competition of moiré superlattices, providing an approach toward engineering the band structure and its topology for a novel quantum material platform with designer electrical and optical properties.

We thank A. Chubukov and B. Shklovskii for useful discussions. This work was supported by NSF DMREF Award DMR-1922165 and Award DMR-1922172. Z.Z. was supported by a Stanford Science Fellowship. E.K. acknowledges support from Simons Foundation Award No. 896626. Nanofabrication was conducted in the Minnesota Nano Center, which was supported by the National Science Foundation through the National Nanotechnology Coordinated Infrastructure (NNCI) under Award No. ECCS-2025124. Portions of the hexagonal boron nitride material used in this work were provided by K.Wat. and T.T. K.Wat. and T.T. acknowledge support from the JSPS KAKENHI (Grants No. 20H00354, No. 21H05233, and No. 23H02052) and World Premier International Research Center Initiative (WPI), MEXT, Japan.

The supporting data for this article are openly available from the Zenodo archive [44].

*kewang@umn.edu

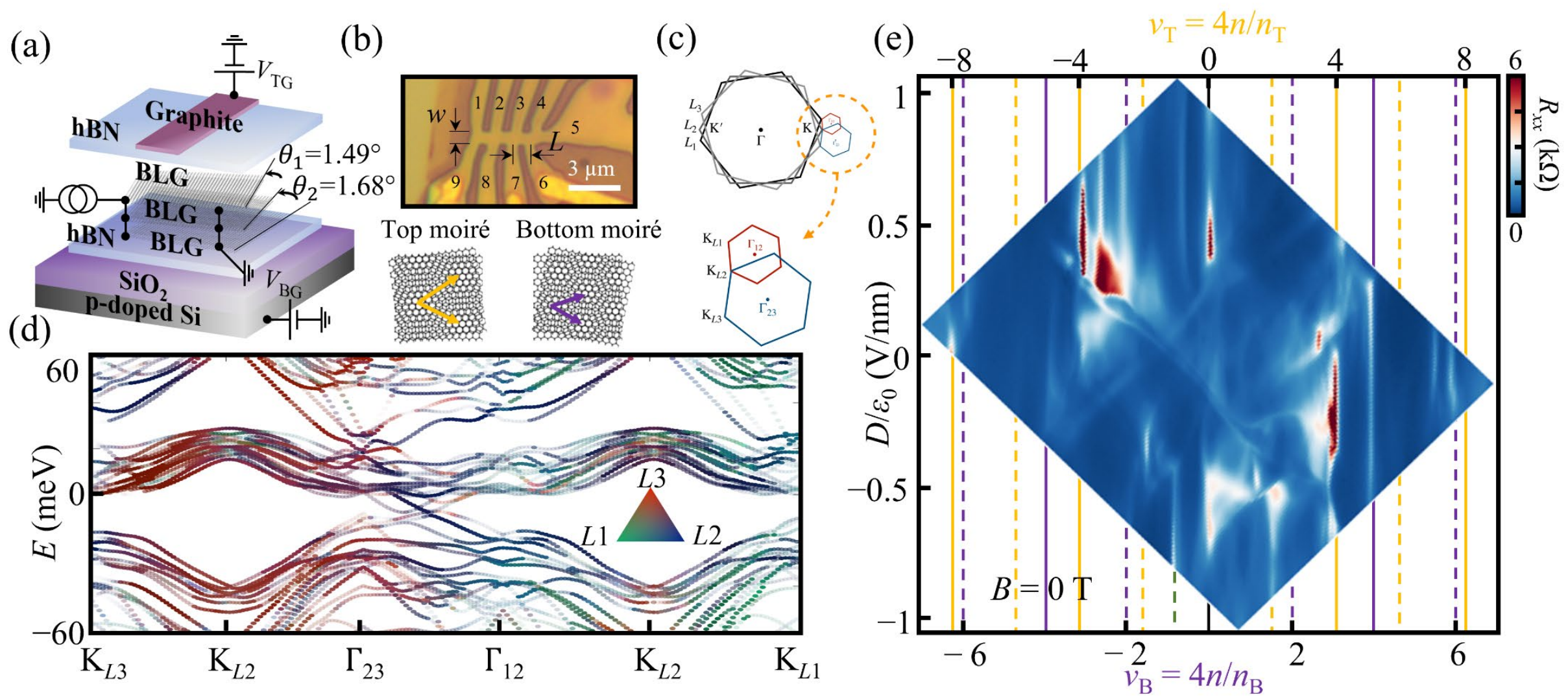

FIG. 1. Gate-tunable insulating states of competing moiré superlattices in tTBLG. (a) Schematic architecture of the dual gated (with applied top, $V_{TG}$, and back, $V_{BG}$, gate voltages) tTBLG stack with the two consecutive twist angles of 1.49° and 1.68° measured under applied current bias. (b) Upper: optical micrograph of the Hall bar shaped tTBLG device with $L$ and $w$ being the length and width of the channel. Lower: schematic images of the coexisting top (bottom) moiré superlattices, with the yellow (purple) arrows indicating the corresponding unequal superlattice vectors. (c) Top: the relative position of the Brillouin zones of each of the three BLGs. Bottom: the alignment of the two mini-Brillouin zones of different sizes corresponding to the top and bottom moirés. (d) Calculated band structure of tTBLG at the twist angles of 1.49° and 1.68° along high symmetry points of the mini-Brillouin zones. Colors indicate the projected weight of each BLG layer. Top layer: green; middle layer: blue; bottom layer: red. (e) The measured four-probe longitudinal resistance $R_{xx}$ as a function of the top (bottom) moiré filling factors and displacement field, $D$. Displacement field asymmetric insulating states are observed at full (solid lines) and half (dashed lines) fillings of the two moiré superlattices.

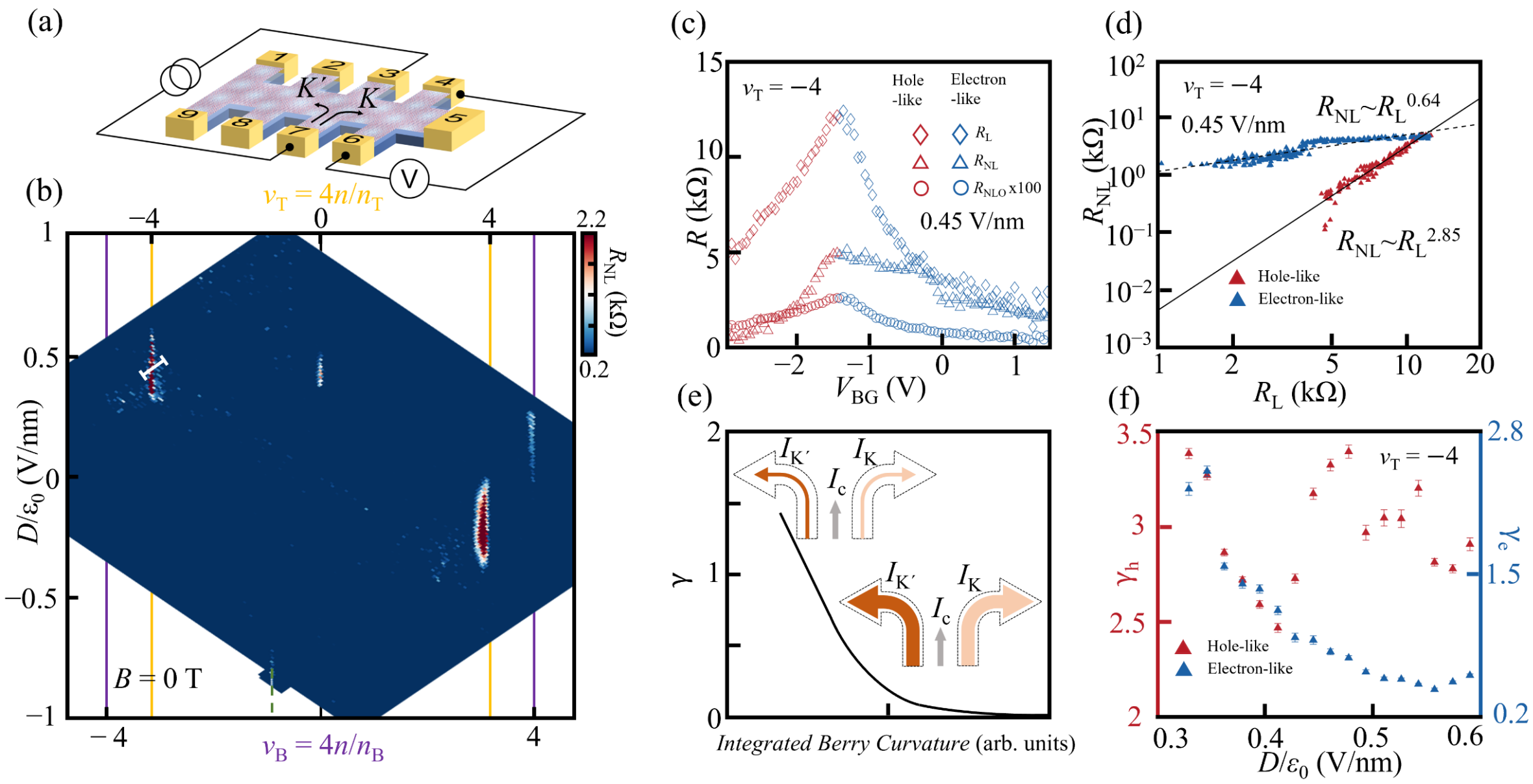

FIG. 2. Atomic lattice reconstruction enhanced moiré Berry curvature hot spot. (a) Nonlocal signal measurement configuration with currents from the different valleys bending in the opposite directions resulting in a transverse valley current. (b) Measured nonlocal resistance $R_{NL} = V_{64}/I_{73}$ (ratio between the measured nonlocal voltage and driven current) as a function of $D$ and $n$. Five peaks are visible corresponding to different Berry curvature hot spots. The white bar represents the horizontal range of the resistance cuts in Fig. 2(c). (c) Nonlocal ($R_{NL}$), local ($R_L$), and nonlocal Ohmic ($R_{NLO}$) resistances due to the stray charge currents near $\nu_T = -4$ and at $D/\varepsilon_0 = +0.45$ V/nm. (d) $R_{NL}$ versus $R_L$ for hole-like (red) and electron-like (blue) carriers near $\nu_T = -4$ at $D/\varepsilon_0 = +0.45$ V/nm. The solid and dashed lines correspond to power law fits for the hole- and electron-like carriers. (e) Qualitative evolution of the power law dependence $R_{NL} \sim R_L^{\gamma}$ as a function of magnitude of the integrated Berry curvature. The inset shows the change in the relative magnitudes of the charge ($I_c$) and the corresponding valley currents ($I_K$, $I_{K'}$) as the Berry curvature (and thus, the ratio between the valley Hall and charge conductance) increases. (f) Displacement field dependence of the powers of the $R_{NL} \sim R_L^{\gamma_h}$ ($R_{NL} \sim R_L^{\gamma_e}$) laws for the hole- (electron-) like charge carriers.

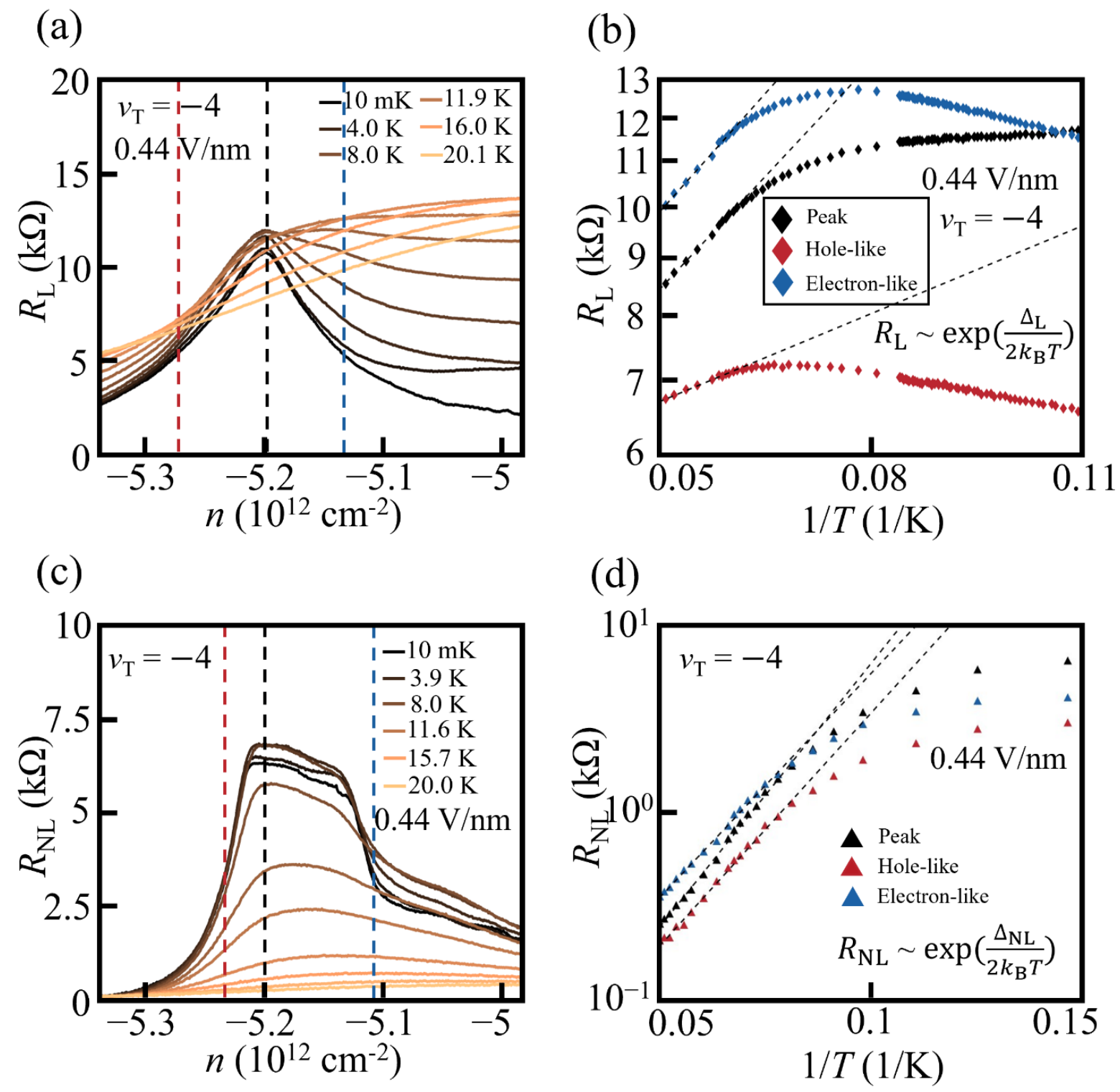

FIG. 3. Temperature dependences of local and nonlocal resistance near moiré Berry curvature hot spot. (a) Thermal activation behavior of $R_L$ near $\nu_T = -4$ at $D/\varepsilon_0 = +0.44$ V/nm. The black dashed line indicates the position of the peak resistance at 10 mK; the blue and red dashed lines trace $R_L$ at half the peak height at 10 mK for the electron- and hole-like charge carriers respectively. Each curve corresponds to a selected temperature increasing from 10 mK to 20 K (indicated by the color) with an increment of $\Delta T \sim 2$ K. (b) Local resistance (black: peak; red: hole-like half-peak; blue: electron-like half-peak) of the $\nu_T = -4$ BI state as a function of inverse temperature $1/T$ at $D/\varepsilon_0 = 0.44$ V/nm. The dashed lines indicate Arrhenius fits of the resistance with thermally activated behavior regime. (c) Same as (a) but for $R_{NL}$ near $\nu_T = -4$ at $D/\varepsilon_0 = +0.44$ V/nm. Similar to (a), the dashed lines indicate $R_{NL}$ at the peak (black) and half-peak (red, blue matching the carrier type) of the resistance curve at 10 mK. Each curve is taken at a different temperature rising from 10mK to 20K with a step of $\Delta T \sim 2$ K. (d) Temperature dependence of the peak (black) and half peak $R_{NL}$ for the electron-like (blue) and hole-like (red) carriers. The dashed lines represent Arrhenius fits at $T > 10$ K.

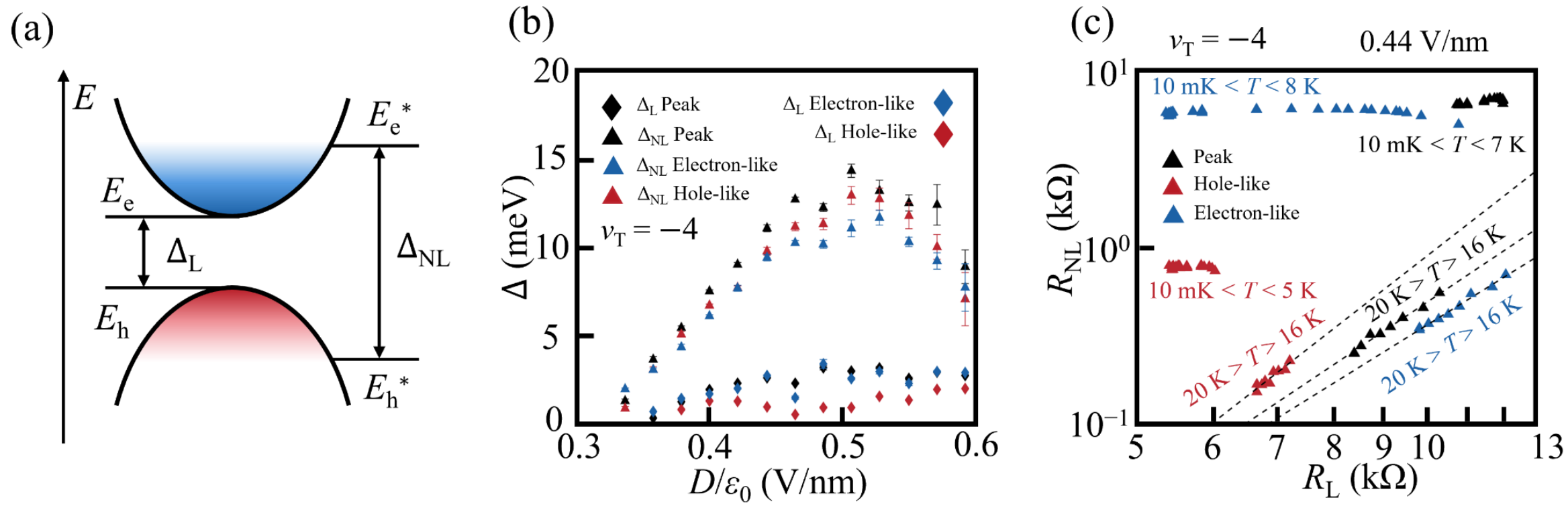

FIG. 4. Characteristic energies of moiré Berry curvature hot spot. (a) Illustration of different energy scales of a moiré Berry curvature hot spot. The local band gap $\Delta_L$ matches the band insulator gap and equals the difference between the conduction and valence band edges energies $E_e$, and $E_h$. The size of the nonlocal band gap $\Delta_{NL} = E_e^* - E_h^*$ determined by the energies $E_e^*(E_h^*)$ above (below) which the Berry curvature labeled by blue (red) vanishes in the electron- (hole-) band. (b) The size of the local ($\Delta_L$) and nonlocal ($\Delta_{NL}$) band gaps found from the Arrhenius fits of the peak (black) and half peak (blue and red, matching the carrier type) value of the resistance as a function of $D$. (c) Log-log plot of $R_{NL}$ (nonlocal resistance) dependence on $R_L$ (local resistance) for electron-like (blue) doping, hole-like (red) doping, and in-gap (black) near $\nu_T = -4$ BI at $D/\varepsilon_0 = +0.44$ V/nm. Data points are taken at varying temperatures. At low temperature range below $T$ ~10 K, the ballistic valley conduction channel is saturated, and thermally activated additional carriers are capable of contributing to local transport only. $R_{NL}$ remains constant while $R_L$ is rapidly decreasing with $T$, resulting in power law of $\gamma < 0.5$. At higher temperature ($T > 10$ K), charge carriers are thermally excited away from Berry curvature hot spot. The weaker Berry curvature makes the ballistic valley conducting channel unsaturated, resulting in an observed power law of $\gamma > 3$.

Supplemental Material

# Tunable atomically enhanced moiré Berry curvatures in twisted triple bilayer graphene

Konstantin Davydov[1], Ziyan Zhu[2], Noah Friedman[1], Ethan Gramowski[1], Yaotian Li[1], Jack Tavakley[1], Kenji Watanabe[3], Takashi Taniguchi[4], Mitchell Luskin[5], Efthimios Kaxiras[6,7], Ke Wang[1*]

[1]*School of Physics and Astronomy, University of Minnesota, Minneapolis, Minnesota 55455, USA*
[2]*Stanford Institute for Materials and Energy Sciences, SLAC National Accelerator Laboratory, Menlo Park, California 94025, USA*
[3] *Research Center for Electronic and Optical Materials, National Institute for Materials Science, 1-1 Namiki, Tsukuba 305-0044, Japan*
[4] *Research Center for Materials Nanoarchitectonics, National Institute for Materials Science, 1-1 Namiki, Tsukuba 305-0044, Japan*
[5]*School of Mathematics, University of Minnesota, Minneapolis, Minnesota 55455, USA*
[6]*Department of Physics, Harvard University, Cambridge, Massachusetts 02138, USA*
[7]*John A. Paulson School of Engineering and Applied Sciences, Harvard University, Cambridge, Massachusetts 02138, USA*
**Corresponding author. Email: kewang@umn.edu*

## S1. Sample preparation, device fabrication, and electrical transport measurements

The dual-gated twisted triple bilayer graphene (tTBLG) stack was made using the "cut and stack" method [1]. A single bilayer graphene (BLG) flake was cut into three individual pieces with the same lattice orientation using the cantilever of an XE7 atomic force microscope (AFM) from Park Systems. Using a poly (bisphenol A carbonate) (PC) and polydimethylsiloxane (PDMS) stamp mounted on a glass slide, a top hexagonal boron nitride [2] (hBN) flake, few layer graphite (serving as the top gate) and middle hBN (isolating the graphite gate from tTBLG) were picked up. Next, the three precut pieces of BLG were successively picked up. After each pick up step, the remaining pieces of BLG were rotated at angles of 1.49° and 1.68° in the same direction. Finally, the tTBLG was encapsulated by picking up a bottom hBN, and the complete stack was released onto a $SiO_2$(285 nm)/Si substrate at 180℃. The PC residue on top of the stack was cleaned by sequentially rinsing the chip containing the stack in chloroform, acetone and isopropanol. Afterwards, a bubble free area of the device serving as the active region was found using the AFM, ensuring the high quality of the device. Ohmic contacts [3] to 1D boundaries of the tTBLG were then added using electron-beam lithography, dry etching, and subsequent metal deposition (Cr/Pd/Au, 1 nm /5 nm/>180 nm). Finally, the active region was given a Hall bar shape [Fig. 1(b)] by a round of electron-beam lithography and reactive ion dry etching.

The electrical and magneto-transport measurements were performed in a Bluefors LD250 cryostat at base temperature of $T = 10$ mK and temperatures ranging from 10 mK to 20 K (for T-dependence) with magnetic fields up to 8 T. The four-probe measurements of the device were performed with 0.5, 1, 10, and 200 nA AC current excitations at a frequency of 17.777 Hz. The voltage drops across the device were measured with the help of SR830 lock-in amplifiers (Stanford Research Systems). The top and back gate voltages were controlled by two voltage sources (Yokogawa: Model GS200 and Keithley Instruments: Model 2400 respectively).

## S2. Calculation of displacement field and charge carrier density from top and bottom gate capacitive coupling

The displacement field, $D$, between the graphite top and silicon back gates and the charge carrier density, $n$, are both calculated by estimating the capacitive coupling of each gate to tTBLG. By using a parallel-plate capacitor model, we find $D = (-C_{TG}V_{TG} + C_{BG}V_{BG})/2 + D_0$, $n = (C_{TG}V_{TG} + C_{BG}V_{BG})/e + n_0$, where $V_{TG}(V_{BG})$ are top (back) gate voltages, $C_{TG}(C_{BG})$ are the top(back) gate capacitances per unit area, $D_0$, $n_0$ are the displacement field and charge carrier offsets, $e$ is the elementary charge. The capacitance of the top gate is given by $C_{TG} = \varepsilon_{hBN}/d_t$, where $\varepsilon_{hBN} = 3.76\varepsilon_0$ is the hBN permittivity, $\varepsilon_0$ is the vacuum permittivity, and $d_t = 23$ nm is the hBN thickness separating the top gate from tTBLG. The back gate has the effective capacitance of two capacitors (with separations due to $SiO_2$ and bottom hBN dielectrics) in series found from $1/C_{BG} = 1/C_{SiO_2} + 1/C_b$, where $C_{SiO_2} = \varepsilon_{SiO_2}/d_{SiO2}$ and $C_b = \varepsilon_{hBN}/d_b$ with $\varepsilon_{SiO_2} = 3.9\varepsilon_0$ being the $SiO_2$ permittivity, $d_{SiO_2} = 285$ nm and $d_b = 44$ nm being the $SiO_2$ and bottom hBN thicknesses respectively. The finite offsets, $D_0$, $n_0$, can be explained by Schottky barriers at the layer interfaces or slight intrinsic doping in tTBLG. The size of $n_0 \sim 1\times10^{11}$ cm$^{-2}$ is characterized by the gate voltage needed to compensate for the offset of the charge neutrality peak from the zero voltage. Same voltage gives $D_0/\varepsilon_0$ of the order of ~ 0.01V/nm.

## S3. Valley Hall effect from additional band insulator states

The valley Hall effect can be observed at multiple BI states in the tTBLG device. In addition to the $\nu_T = -4$ and MoM BI states, Berry curvature hot spots are present near the $\nu_T = 4$ and $\nu_B = 4$ fillings [Fig. 2(b)]. At both fillings, nonlocal measurements following the protocol from the main manuscript demonstrate a nonlocal resistance more than one order of magnitude larger than trivial Ohmic contribution due to stray currents [Figs. S1(a) and S1(b)]. Figures S1(c) and S1(d) show the nonlocal resistance, $R_{NL}$, as a function of the local one $R_L$ near the $\nu_T = 4$ and $\nu_B = 4$ fillings respectively. The plotted resistance dependence shows a clear difference between the hole-like (red) and electron-like (blue) carriers changing the type across the BI state. The nonlocal resistance for each carrier type follows a nonlinear power law relationship $R_{NL} \sim R_L^{\gamma}$ with the power $\gamma = \gamma_h$ ($\gamma_e$) extracted from separate fittings for hole- (electron-) like $R_{NL}$. Both powers being sufficiently different from $\gamma = 1$ at varying $D$ [Figs. S1(e) and S1(f)] indicate a non-Ohmic mechanism of nonlocal transport through valley current due to a Berry curvature hot spot. When the ballistic valley-chiral channel is completely saturated, the power $\gamma$ reaches zero. In the opposite scenario, when the valley-chiral channel is established, yet far from saturation, the power is expected to approach $\gamma \rightarrow 3$.

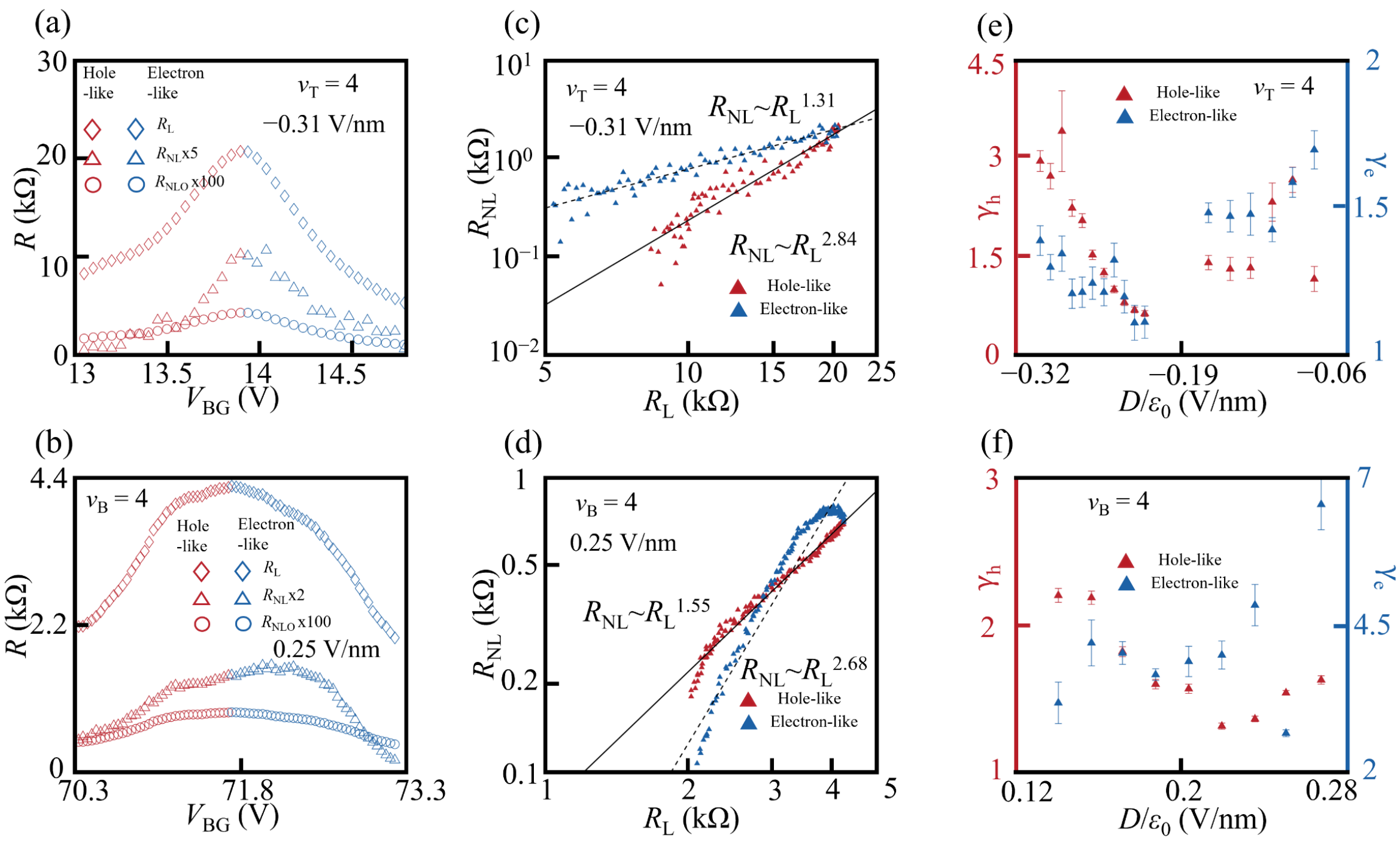

FIG. S1. Valley Hall effect at $\nu_T = 4$ and $\nu_B = 4$ in tTBLG. (a) and (b) Back gate voltage dependance of the local ($R_L$), nonlocal ($R_{NL}$) and nonlocal Ohmic resistance ($R_{NLO}$) near the $\nu_T = 4$ and $\nu_B = 4$ BI states at $D/\varepsilon_0 = -0.31$ V/nm and $D/\varepsilon_0 = 0.25$ V/nm respectively. The color highlights the types of carriers participating in electrical transport. (c) and (d) The dependence of $R_{NL}$ on $R_L$ for hole-like (red) and electron-like (blue) carriers at $\nu_T = 4$ and $\nu_B = 4$ and the same $D$ as in (a) and (b). For each carrier type, the trendlines show a power law fit $R_{NL} \sim R_L^{\gamma}$. (e) and (f) The extracted power $\gamma = \gamma_h$ ($\gamma_e$) for the hole- (electron-) like carriers as a function of displacement field at $\nu_T = 4$ and $\nu_B = 4$.

## S4. Temperature dependance of band insulator states in tTBLG

By following the fitting procedure described in the main text, we extract the band and nonlocal gaps of the $\nu_T = 4$, $\nu_B = 4$ BI states from the temperature dependence of local and nonlocal resistances. Figs. S2(a) and S2(b) summarize the sizes of the band insulator gaps ($\Delta_L$, diamonds) as a function of displacement field $D$ comparing them with the nonlocal gaps ($\Delta_{NL}$, triangles) found from the thermal activation behavior of the nonlocal resistance $R_{NL}$ as discussed in the main manuscript. The nonlocal gaps are consistently larger than the band insulator ones confirming the presence of a Berry curvature hot spot near the band edges. As the temperature increases, the charge carriers are thermally excited between the conduction and valence bands which results in diminishing $R_L$. However, the larger $\Delta_{NL}$ compared to $\Delta_L$, indicates that additional energy is needed to push the carriers away from the Berry curvature hot spot.

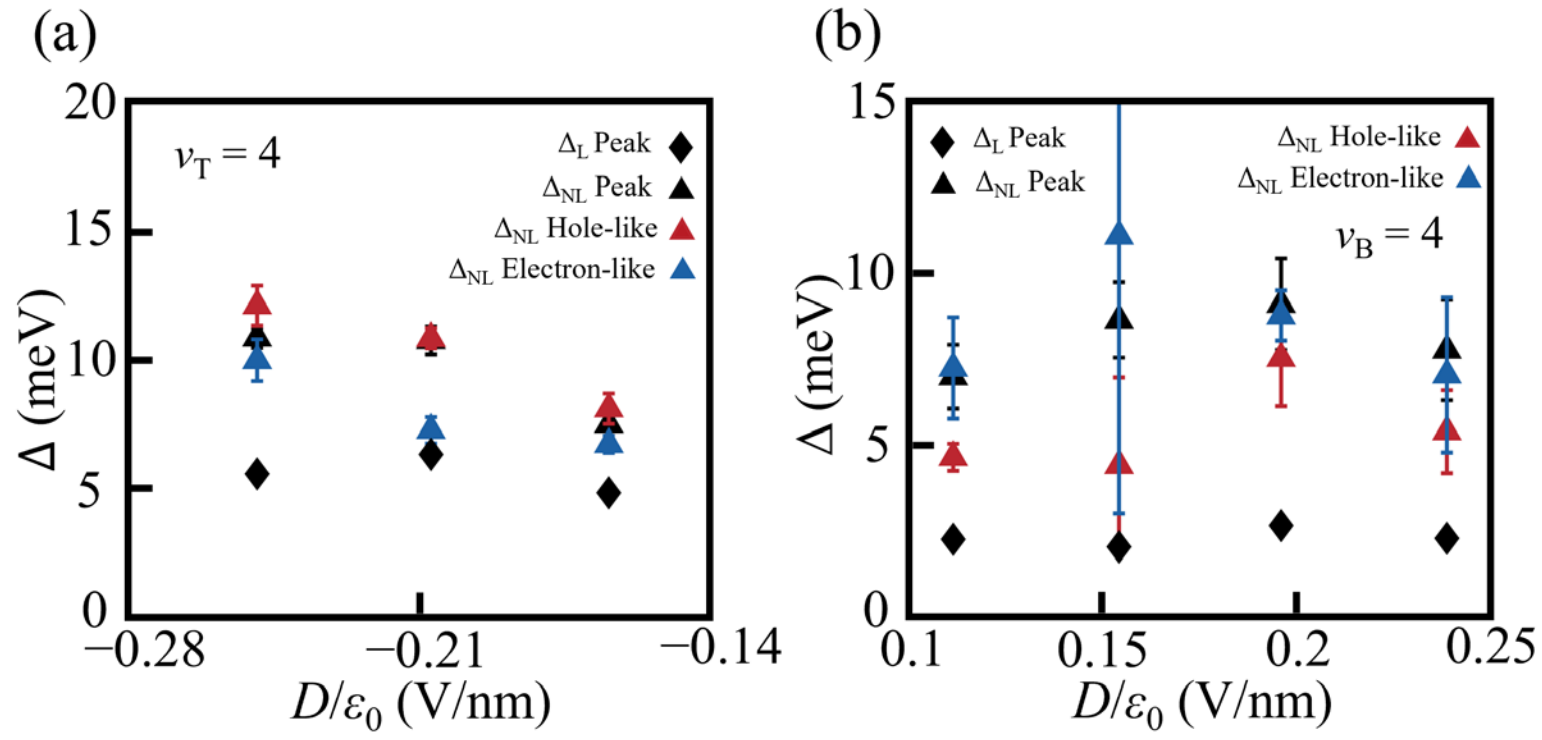

FIG. S2. Temperature dependence of the $\nu_T = 4$ and $\nu_B = 4$ band insulator states in tTBLG. (a) and (b) Displacement field dependence of band insulator $\Delta_L$ (diamonds) and nonlocal band gap $\Delta_{NL}$ (triangles) found from Arrhenius fits of the $\nu_T = 4$ and $\nu_B = 4$ BI resistance peak (black) and half peak (red and blue, with the colors corresponding to hole-like and electron-like carrier types).

## S5. Results from a different region of tTBLG device

Band insulator valley Hall effect was reproduced in a different region of the same tTBLG device discussed in the main manuscript (Device 1). The main manuscript discusses the measured local resistance, $R_{xx}$, in Region 1 between Contact 3 and Contact 4 of the device shown in Fig. 1(b). Similarly, under the same driven current (Contact 5 serving as the source; Contacts 1, 9 serving as the drain), the four-probe longitudinal resistance, $R_L = R_{xx}$, between Contact 2 and Contact 3 corresponding to Region 2 of the same device can be measured. In this region, in addition to correlated insulator states at moiré half filling (Fig. S3), displacement field tunable band insulator states are observed at four charge carriers per top (bottom) moiré unit cell corresponding to twist angles of 1.48°±0.02° (1.66°±0.01°).

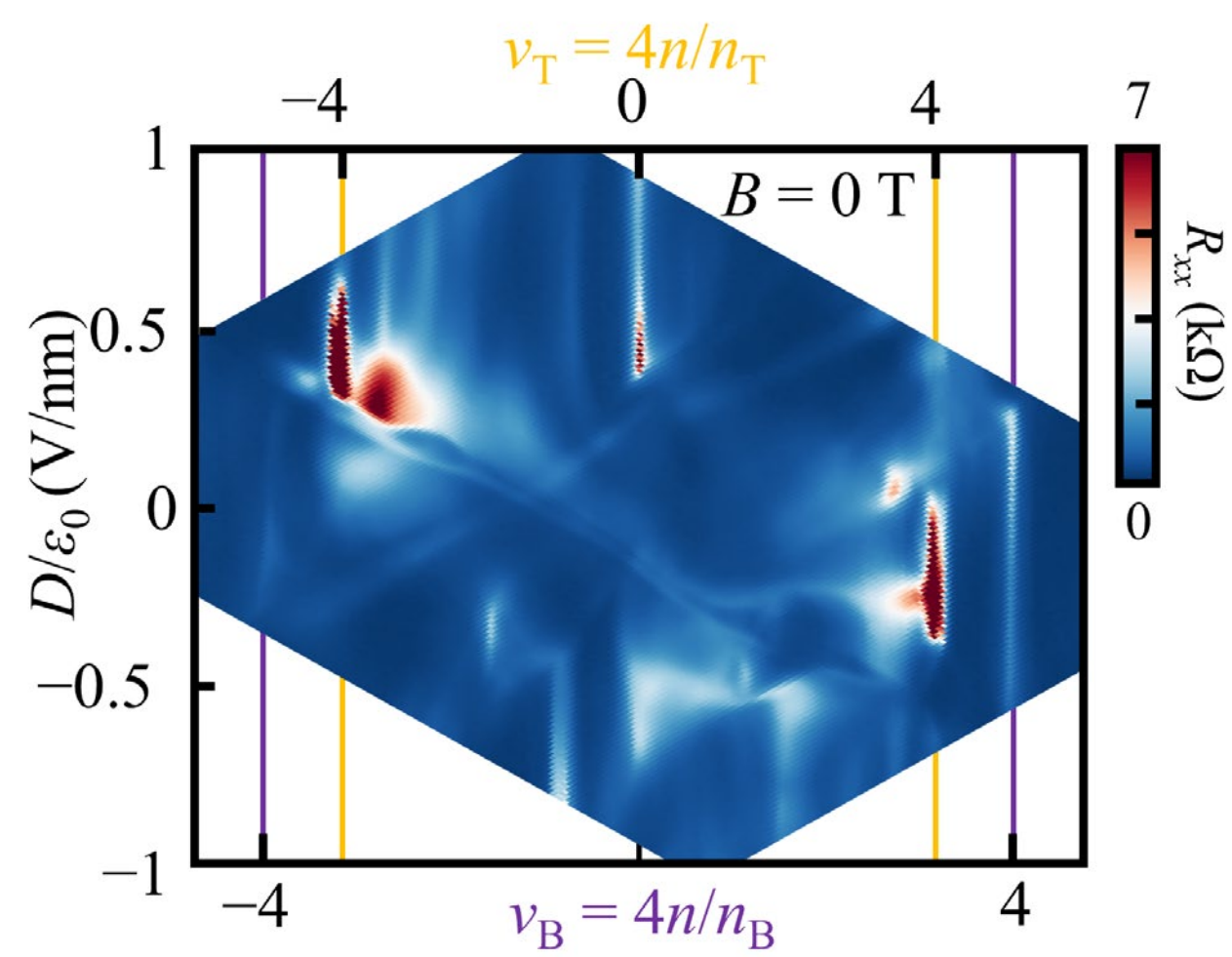

FIG. S3. Local transport in Region 2 of Device 1. Longitudinal resistance $R_{xx}$ between Contact 2 and Contact 3 as a function of charge carrier density $n$ and displacement field $D$. Resistance peaks of band insulator states (highlighted by solid lines) appearing at four charge carriers per top (yellow) and bottom (purple) moiré unit cell.

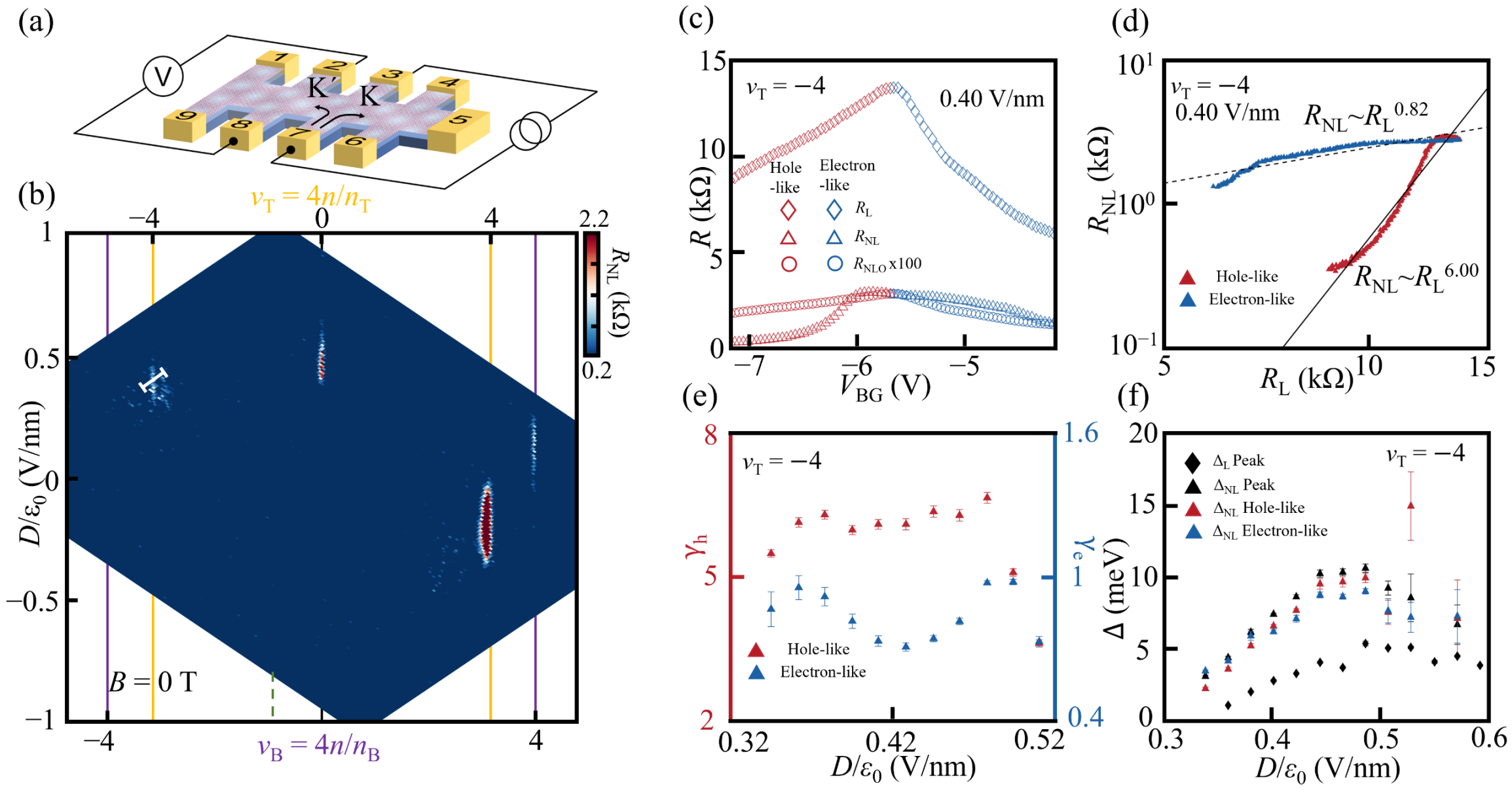

FIG. S4. Valley Hall effect in Region 2 of the tTBLG device. (a) Measurement configuration for studying nonlocal transport in Region 2 of the tTBLG device. (b) Nonlocal resistance $R_{NL} = V_{82}/I_{73}$ found from the driven current and measured voltage in the configuration of (a) at varying $D$ and $n$. The resistance peaks corresponding to the Berry curvature hot spots at $\nu_T = \pm 4$, $\nu_B = 4$ are reproduced in Region 2. The white bar represents the horizontal range in (c). (c) Nonlocal ($R_{NL}$), local ($R_L$), and calculated nonlocal Ohmic resistance ($R_{NLO}$) contributing to nonlocal transport of the $\nu_T = -4$ BI at $D/\varepsilon_0 = +0.40$ V/nm. (d) $R_{NL}$ as a function of $R_L$ for hole-like (red) and electron-like (blue) carriers across the $\nu_T = -4$ BI band gap at $D/\varepsilon_0 = +0.40$ V/nm. The dependence is consistent with power law fits for the hole-like (solid line) and electron-like (dashed line) carriers. (e) Displacement field dependence of the powers extracted from the $R_{NL}{\sim}R_L^{\gamma_h}$ ($R_{NL}{\sim}R_L^{\gamma_e}$) fits for the hole- (electron-) like charge carriers. (f) Estimated from the Arrhenius fits, the local $\Delta_L$ nonlocal $\Delta_{NL}$ bandgaps at the BI peak (black) and half peak (blue and red, corresponding to the electron- and hole-like carriers) value of the resistance as a function of $D$.

By driving current, $I_{73}$, between Contact 7 and Contact 3 and measuring voltage, $V_{82}$, between Contact 8 and Contact 2 [Fig. 1(b) and Fig. S4(a)], a nonlocal resistance, $R_{NL} = V_{82}/I_{73}$, in Region 2 can be found as a function of $D$ and $n$ [Fig. S4(b)]. Similar to Region 1, a finite nonlocal resistance is measured at $\nu_T = \pm 4$, $\nu_B = 4$, indicating the presence of Berry curvature hot spots. Near $\nu_T = -4$, the measured $R_{NL}$ [Fig. S4(c)] is two orders of magnitude larger than the trivial Ohmic resistance from the stray charge currents. This points to the presence of a transverse valley current responsible for the increased nonlocal resistance. The plotted $R_{NL}$ as a function of $R_L$ [Fig. S4(d)] reveals two separate nonlinear power law dependences, $R_{NL} \sim R_L^{\gamma}$, characteristic to a valley Hall effect, for the hole-like ($\gamma = \gamma_h$) and electron-like ($\gamma = \gamma_e$) charge carrier types changing across the $\nu_T = -4$ insulator band gap. Changing $D$ has no consistent effect on the form of the power law dependence [Fig. S4(e)]. However, over the entire range of the displacement field, both powers differ from $\gamma = 1$ excluding the possibility of trivial Ohmic transport. Found from the resistance behavior during thermal activation, the nonlocal bandgap, $\Delta_{NL}$, turns out to be higher than $\Delta_L$ over the entire range of displacement fields at which the nonlocal signal exists [Fig. S4(f)]. This is consistent with the Berry curvature hot spot model depicted in main manuscript Fig. 4(a).

The main experimental observations of atomically enhanced Berry curvature have been reproduced in a different region of tTBLG, suggesting the robustness of the observed emergent quantum phenomena. The differences in quantitative details (i.e., the extracted power law dependence) further supports that the Berry curvature of reconstructed bands is extremely sensitive to microscopic details of reconstructed local atomic landscape, which can realistically vary with the presence of even slight twist angle inhomogeneity.

## S6. Calculation of the electronic structure

To obtain the electronic structure of tTBLG, we follow the momentum-space model in [4] with modified hopping parameters. The Hamiltonian in momentum space can then be formally written as

$$H_K(\boldsymbol{q}) = \begin{bmatrix} H^1(\boldsymbol{q}) + \delta_1 \mathbb{I}_{2\times 2} & T^{12} & 0 \\ T^{12\dagger} & H^2(\boldsymbol{q}) + \delta_2 \mathbb{I}_{2\times 2} & T^{23} \\ 0 & T^{23\dagger} & H^3(\boldsymbol{q}) + \delta_3 \mathbb{I}_{2\times 2} \end{bmatrix}. \quad (S2)$$

The diagonal blocks describe the low-energy Hamiltonian of the Bernal bilayer graphene with $\delta_\ell$ representing the onsite potential due to the applied vertical displacement field. The intralayer parameters are taken from [5]. The off-diagonal terms describe the interlayer interaction between adjacent Bernal bilayers. More explicitly, the intralayer terms in the $(A_1, B_1, A_2, B_2)$ basis can be expressed as:

$$H^\ell(\boldsymbol{q}) = \begin{bmatrix} H_D^{\ell,t}(\boldsymbol{q}) & g^\dagger(\boldsymbol{q}) \\ g(\boldsymbol{q}) & H_D^{\ell,b}(\boldsymbol{q}) \end{bmatrix}, \quad (S3)$$

where the superscript $t/b$ denotes the top and bottom Bernal bilayer graphene. $H_D$ is the rotated Dirac Hamiltonian for monolayer graphene.

$$H^{\ell,t/b}(\boldsymbol{q}) = \begin{bmatrix} \Delta_{\mathrm{t}} & \hbar v_F e^{i\theta_\ell} q_+ \\ \hbar v_F e^{i\theta_\ell} q_- & \Delta_{\mathrm{b}} \end{bmatrix}, \quad (S4)$$

where $\theta_1 = \theta_{12}$, $\theta_2 = 0$, and $\theta_3 = \theta_{23}$, and $v_F = 0.8 \times 10^6$ m/s is the monolayer graphene Fermi velocity, $q_\pm = q_x \pm q_y$. The diagonal $\Delta_{\mathrm{t/b}}$ represents the on-site potential of dimer sites with respect to nondimer sites, and $\Delta_{\mathrm{t/b}} = 0.050$ eV is the neighboring layer is not vacuum and 0 otherwise. $g(\boldsymbol{k})$ is the parabolic part of the band structure:

$$g(\boldsymbol{q}) = \begin{bmatrix} \hbar v_4\, q_+ & \gamma_1 \\ \hbar v_3 q_- & \hbar v_4 k_+ \end{bmatrix}, \quad (S5)$$

where $\gamma_1 = 0.4$ eV is the hopping between dimer sites, $\gamma_3 = 0.32$ eV, $\gamma_4 = 0.044$ eV and $v_i = \frac{\sqrt{3}\gamma_i a}{2\hbar}$ where $a$ is the monolayer graphene lattice constant.

For the interlayer coupling, we keep the nearest neighbor coupling in momentum space

$$T_{\alpha\beta}^{ij}\left(\boldsymbol{q}^{(i)}, \boldsymbol{q}^{(j)}\right) = \sum_{n=1}^{3} T_{\alpha\beta}^{\boldsymbol{q}_n^{ij}}\, \delta_{\boldsymbol{q}^{(i)} - \boldsymbol{q}^{(j)}, -\boldsymbol{q}_n^{ij}}, \quad (S6)$$

where $\boldsymbol{q}_1^{ij} = K_{\mathrm{L_i}} - K_{\mathrm{L_j}}$, $\boldsymbol{q}_2^{ij} = \mathcal{R}^{-1}\left(\frac{2\pi}{3}\right)\boldsymbol{q}_1^{ij}$, $\boldsymbol{q}_3^{ij} = \mathcal{R}\left(\frac{2\pi}{3}\right)\boldsymbol{q}_1^{ij}$ and $\mathcal{R}(\theta)$ is the counterclockwise rotation matrix by $\theta$. We take into account the out-of-plane relaxation by letting

$t_{AA}^{ij} = t_{BB}^{ij} = \omega_0 = 0.07$ eV and $t_{AB}^{ij} = t_{BA}^{ij} = \omega_1 = 0.11$ eV due to the strengthened interaction between AB/BA sites from relaxation.

$$T^{q_1^{ij}} = \begin{bmatrix} \omega_0 & \omega_1 \\ \omega_1 & \omega_0 \end{bmatrix}, T^{q_2^{ij}} = \begin{bmatrix} \omega_0 & \omega_1\bar{\phi} \\ \omega_1\phi & \omega_0 \end{bmatrix}, T^{q_3^{ij}} = \begin{bmatrix} \omega_0 & \omega_1\phi \\ \omega_1\bar{\phi} & \omega_0 \end{bmatrix}, \quad (S7)$$

where $\phi = e^{\frac{i2\pi}{3}}$ and $\bar{\phi} = e^{-\frac{i2\pi}{3}}$.

## S7. Spatial dependence of valley currents in tTBLG

By following the previously-introduced approaches [6–12], we measure the nonlocal signal $R_{\mathrm{NL}}$ at voltage probes located at varying distance $L$ from the applied current [Fig. S5 and Fig. 1(b)]. In this configuration, at an electric field $E$ driving the charge current, the Berry curvature hot spot yields a non-zero transverse valley current density (valley Hall effect), $J_{\mathrm{v}} = \sigma_{xy}^{v}E$. At a band insulator state, the transverse valley Hall conductivity $\sigma_{xy}^{v}$ is expected to be $Me^2/h$, where $M$ is the total Berry flux of the occupied states below the Fermi energy. Subsequently, the inverse valley Hall effect is responsible for generating the nonlocal signal $R_{\mathrm{NL}}$ decaying with $L$ [Fig. S5(b)]. Following the previously established methods [6,8–10], we fit the length dependence of $R_{\mathrm{NL}}$ at the $\nu_{\mathrm{T}} = -4$ [Fig. S5(b)] BI state with the model expression

$$R_{\mathrm{NL}} = \left(\frac{w}{2\xi}\right)\left(\sigma_{xy}^{v}\right)^2 \rho_{xx}^3 \exp\left(-L/\xi\right), \quad (S8)$$

where $w$ is the width of the valley channel, $\xi$ is the valley decay length, and $\rho_{xx}$ is the longitudinal resistivity. For $\nu_{\mathrm{T}} = -4$ BI we found $\xi$ to be 4±2 $\mu$m, a change much slower than the decay due to diffusive Ohmic transport, and consistent with the expected inter-valley scattering lengths. For the $\nu_{\mathrm{T}} = -4$ BI state, $\sigma_{xy}^{v}$ is found to be ~$3e^2/h$ confirming the Berry curvature hot spot being the origin of the nonlocal signal.

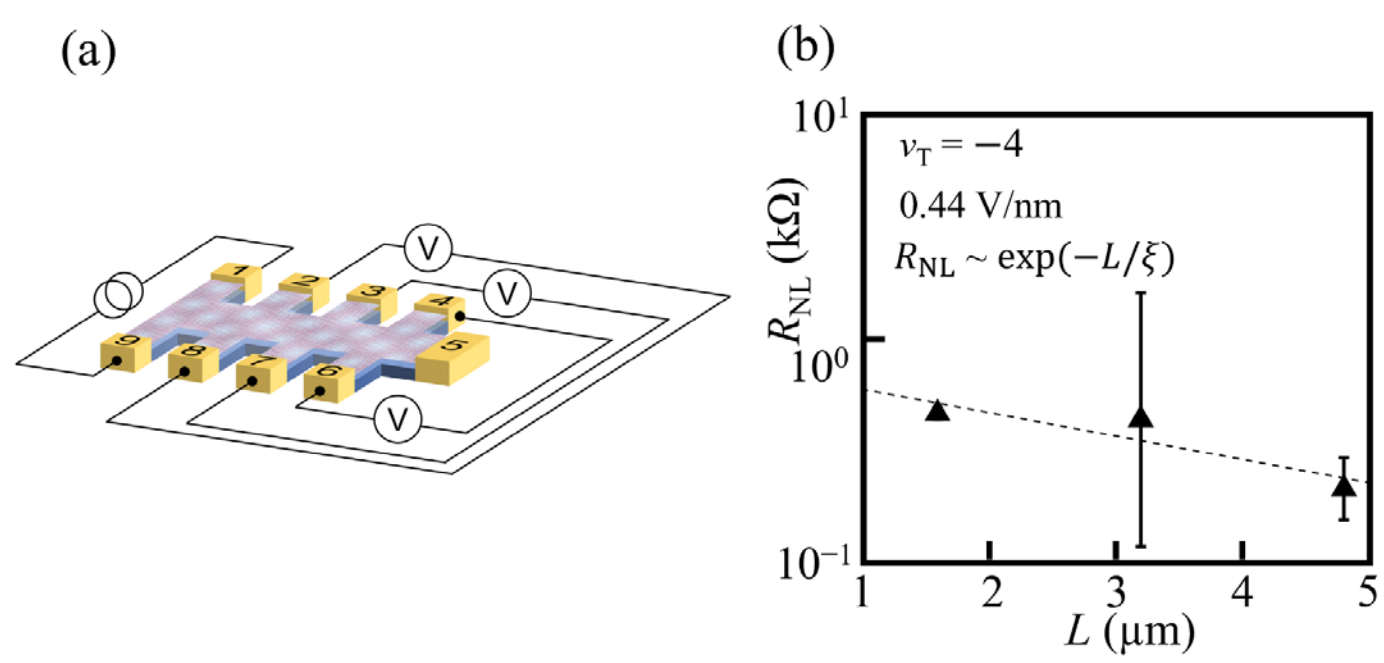

FIG. S5. Spatial dependence of valley Hall effect. (a) Nonlocal transport measurement configuration with non-local voltages simultaneously measured at different distances from the driven current. (b) $R_{\mathrm{NL}}$ calculated from the measured nonlocal voltages as a function of $L$, the distance from the driven current, at $\nu_{\mathrm{T}} = -4$ ($D/\varepsilon_0 = +0.44$ V/nm) BI state. The dashed lines represent a data fit with Equation ($S8$).

## S8. Valley currents measured in additional device

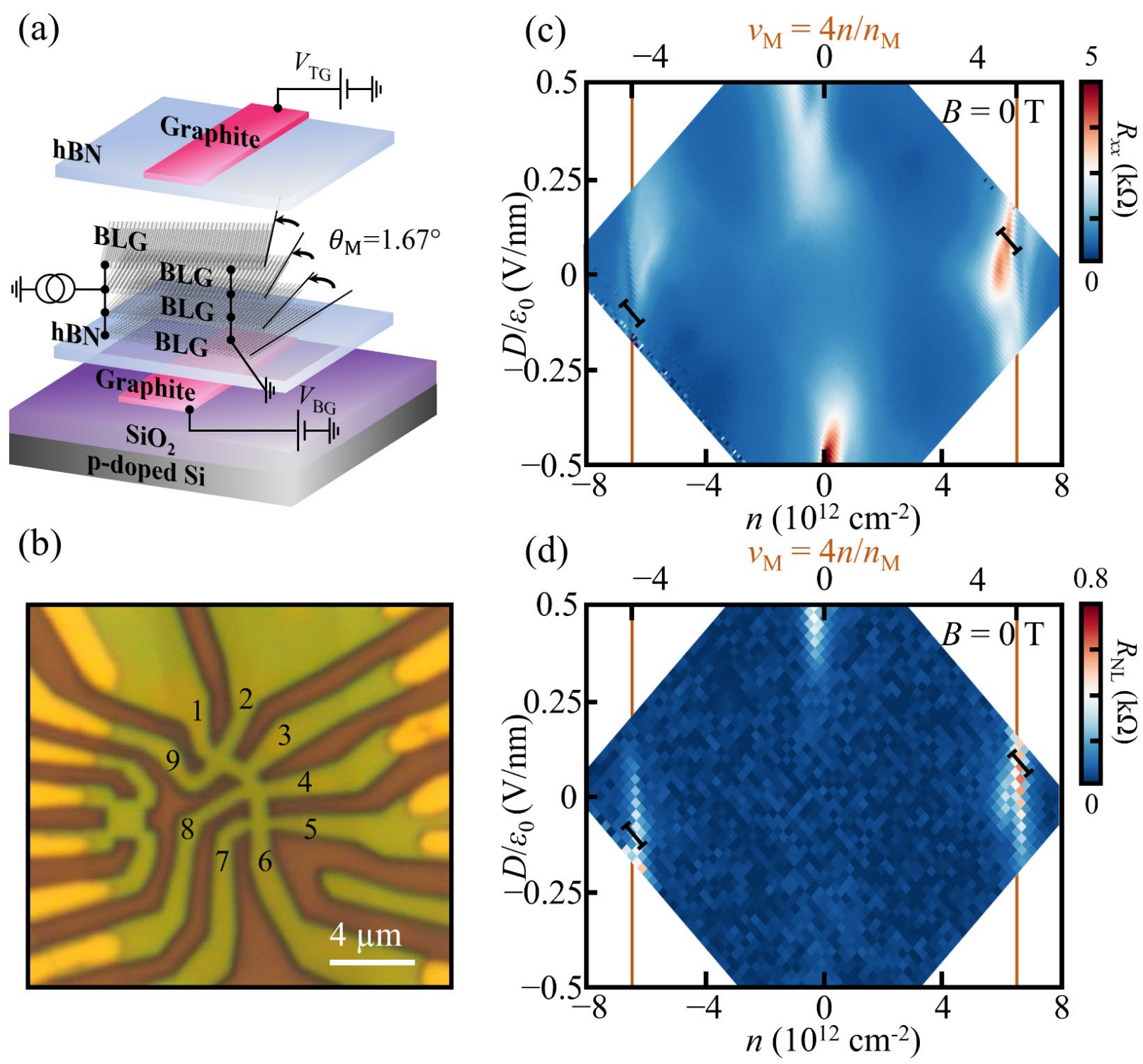

FIG. S6. Results from second device. (a) Schematic structure of the second dual gated twisted multilayer device consisting of four layers of BLG consecutively twisted in the same direction. The voltage applied to the top ($V_{TG}$) and bottom ($V_{BG}$) graphite gates controls the charge carrier density and displacement field. (b) Optical micrograph of the second device with the numbers labeling contacts to the twisted multilayer region. (c) The measured four-probe resistance $R_{xx}$ between Contacts 4 and 5 as a function of total charge carrier density $n$, middle moiré filling factor $\nu_M$ and out-of-plane displacement field $D$. The current is driven between Contacts 1 and 6. The color plots in (c) and (d) have black bars indicating the horizontal top gate voltage ranges in Figs. S7(a) and S7(b). (d) Same as (c) but for the measured nonlocal resistance between Contacts 4 and 8 when a current is driven between Contacts 5 and 7.

Displacement field-tunable valley currents were also observed in an additional twisted multilayer system [Device 2, Figs. S6(a) and S6(b)]. The second device consists of four layers of BLG consecutively twisted with respect to each other [Fig. S6(a)]. The device was fabricated by following a similar protocol introduced in Section S1. However, an additional graphite bottom gate was picked up before the entire stack was released onto a $SiO_2$ substrate. The displacement field and charge carrier densities were calculated based on a parallel-place capacitor model from Section S2: $D = (-C_{TG}V_{TG} + C_{BG}V_{BG})/2 + D_0$, $n = (C_{TG}V_{TG} + C_{BG}V_{BG})/e + n_0$, where, for Device 2, $C_{TG} = \varepsilon_{hBN}/d_t$ ($C_{BG} = \varepsilon_{hBN}/d_b$) is the capacitance per unit area of the top (bottom) graphite gate. The thickness of the hBN flake separating the top (bottom) graphite gate from the twisted multilayer is respectively $d_t = 47$ nm ($d_b = 99$ nm). As in Section S2, we estimate the charge carrier density and displacement field offsets to be $D_0/\varepsilon_0$ ~0.02 V/nm and $n_0 \sim 2\times10^{11}$ cm$^{-2}$ respectively.

We first characterize Device 2 by driving a current between Contact 1 and Contact 6 [Fig. S6(b)] and measuring the local four-probe resistance, $R_L = R_{xx}$, between Contact 4 and Contact 5 as a function of $n$ and $D$ [Fig. S6(c)]. Near $n \sim \pm 6.5\times10^{12}$ cm$^{-2}$, we observe resistance peaks from moiré band insulator states corresponding to a twist angle of $\theta_M = 1.67°\pm0.05°$. Their dependence on the displacement field points to the origin of the BI being from the moiré between the two middle BLG layers. Namely, applying a small either positive or negative displacement field pushes the charges away from the middle moiré interface. Thus, the middle moiré BI states are effectively shortened by either the topmost or bottommost conductive BLG which results in a decreasing $R_{xx}$.

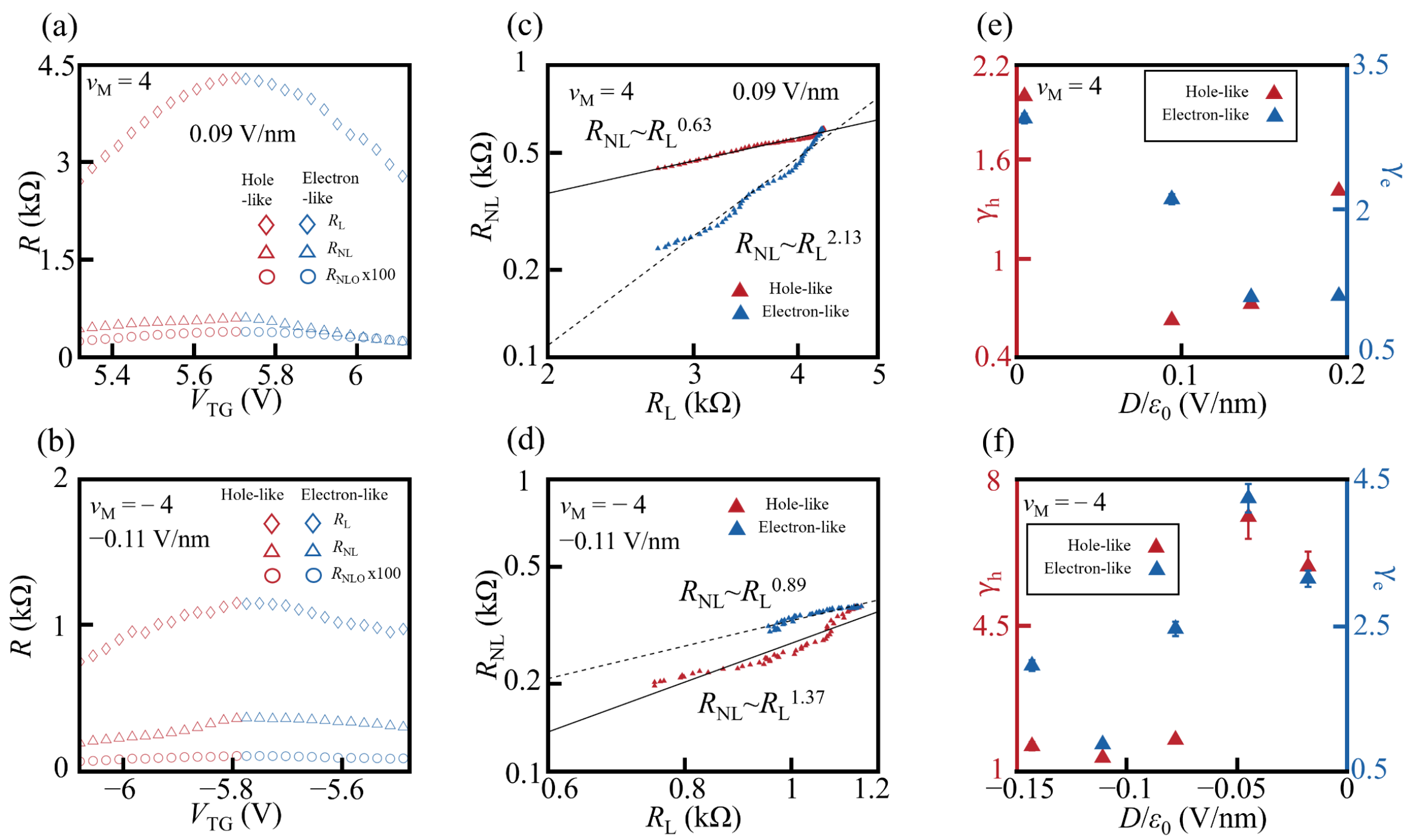

FIG. S7. Valley Hall effect in second device. (a) and (b) The dependance of the local ($R_L$), nonlocal ($R_{NL}$) and nonlocal Ohmic resistances ($R_{NLO}$) on the top graphite gate voltage ($V_{TG}$) near $\nu_M = 4$ and $\nu_M = -4$ BI states at $D/\varepsilon_0 = 0.09$ V/nm and $D/\varepsilon_0 = -0.11$ V/nm correspondingly. The color indicates the types of carriers involved in electrical transport. (c) and (d) $R_{NL}$ as a function of $R_L$ for hole-like (red) and electron-like (blue) carriers near $\nu_M = \pm4$ and the displacement fields as in (a) and (b). For each carrier type, the solid and dashed lines are power law fits $R_{NL}{\sim}R_L^{\gamma}$. (e) and (f) The fitted curve exponents $\gamma = \gamma_h$ ($\gamma_e$) for the hole- (electron-) like carriers as a function of displacement field near $\nu_M = \pm4$.

At middle moiré fillings of $\nu_M = \pm4$ in Device 2, we observe a substantial nonlocal resistance $R_{NL} = V_{48}/I_{57}$ [Fig. S6(d)], when a current is driven from Contact 5 to Contact 7, and the voltage is measured between Contact 4 and Contact 8. The measured nonlocal resistance at $\nu_M = \pm4$ is more than two orders of magnitude larger [Figs. S7(a) and S7(b)], than the expected contribution from nonlocal Ohmic currents $R_{NLO} = R_L(w/L)\exp(-L/\lambda)$, where $\lambda = w/\pi$; $w = 0.8$ µm, $L = 1.6$ µm are the width and length of the channel. Therefore, the measured nonlocal signal cannot be due to stray Ohmic currents and should come from a Berry curvature hot spot. To confirm this, we study the dependence of $R_{NL}$ on $R_L$ near the BI peak for electron-(Fig. S7, blue), and hole-like (Fig. S7, red) carriers controlled by the graphite top gate voltage. For a selected carrier type (either hole- or electron-like) near the $\nu_M = \pm 4$ BI states, the log-log plots [Figs. S7(c) and S7(d)] of $R_{NL}$ as a function of $R_L$ exhibit a clear power law dependence fitted by a trendline $R_{NL} \sim R_L^{\gamma}$. For the fitted curves plotted as solid (dashed) lines in Figs. S7(c) and S7(d) corresponding to the hole- (electron-like) carriers, we extract the exponents $\gamma_h(\gamma_e)$ whose values indicate the strength of the valley Hall effect.  The latter can be tuned by applied out-of-plane displacement field [Figs. S7(e) and S7(f)]. Near $\nu_M = 4(-4)$, both carrier type exponents exhibit a trend of having the lowest values near $D_0/\varepsilon_0 \sim 0.1(-0.1)$ V/nm, being a signature of $\gamma$ decreasing toward zero due to saturation of valley conducting channel. Specifically, for hole-like carriers near $\nu_M = 4$ at $D_0/\varepsilon_0 \sim 0.09$ V/nm, the exponent of the scaling relation $R_{NL} \sim R_L^{\gamma}$ reaches $\gamma_h = 0.63$ suggesting a significant reinforcement of the Berry curvature hot spot. This can potentially be explained by two competing inter-moiré interactions between the middle and topmost moiré interfaces and between the middle and bottommost moirés. At $|D_0/\varepsilon_0| < 0.1$ V/nm, both the topmost and the bottommost moirés are depleted of charge carriers thus equally affecting the middle moiré interface and resulting in further amplification of the valley currents.